\documentclass[pra,twocolumn,showpacs,preprintnumbers,amsmath,amssymb,superscriptaddress]{revtex4}
\usepackage{graphicx,amsmath,amssymb,amsfonts,latexsym,color,dcolumn,bm,epsfig,subfigure}
\usepackage[latin1]{inputenc}

\def\bbm[#1]{\mbox{\boldmath $#1$}}
\newcommand{\ket}[1]{\displaystyle{|#1\rangle}}
\newcommand{\bra}[1]{\displaystyle{\langle #1|}}

\newcommand{\TE}{\text{TE}}
\newcommand{\TM}{\text{TM}}

\begin{document}
\title{Non equilibrium dissipation-driven steady many-body entanglement}

\author{Bruno Bellomo}\affiliation{Universit\'{e} Montpellier 2, Laboratoire Charles Coulomb UMR 5221 - F-34095, Montpellier, France}\affiliation{CNRS, Laboratoire Charles Coulomb UMR 5221 - F-34095, Montpellier, France}

\author{Mauro Antezza}\affiliation{Universit\'{e} Montpellier 2, Laboratoire Charles Coulomb UMR 5221 - F-34095, Montpellier, France}\affiliation{CNRS, Laboratoire Charles Coulomb UMR 5221 - F-34095, Montpellier, France}\affiliation{Institut Universitaire de France - 103, bd Saint-Michel
F-75005 Paris, France, EU}

\begin{abstract}We study an ensemble of two-level quantum systems (qubits) interacting with a common electromagnetic field in proximity of a dielectric slab whose temperature is held different from that of some far surrounding walls. We show that the dissipative dynamics of the qubits driven by this stationary and out of thermal equilibrium (OTE) field, allows the production of steady many-body entangled states, differently from the case at thermal equilibrium where steady states are always non-entangled. By studying up to ten qubits, we point out the  role of symmetry in the entanglement production, which is exalted in the case of permutationally invariant configurations. In the case of three qubits, we find a strong dependence of  tripartite entanglement on the spatial disposition of the qubits, and in the case of six qubits, we find several highly entangled bipartitions where entanglement  can, remarkably, survive for large qubit-qubit distances  up to 100 $\mu$m.

\end{abstract}


\maketitle

\section{Introduction}

Entanglement is a central notion in quantum mechanics being considered the most nonclassical manifestation of quantum formalism \cite{Horodecki09}.
It has been recognized as a new resource for tasks that cannot be performed by means of classical ones and it  can be manipulated,
broadcast, controlled and distributed \cite{BookNielsen,Horodecki09}. In many-body quantum  systems different types of multipartite entanglement can be present, such as bipartite and genuine entanglement \cite{Guhne2011}, this last being viewed as a resource for quantum information processing, as for example in measurement-based quantum computation \cite{Briegel2009}.

Due to  the rapid development of quantum experiments, it is now possible to create highly entangled multiqubit
states using photons \cite{Wieczorek2008}, trapped ions \cite{Haffner2005}, and cold atoms \cite{Mandel2003}.
In particular, the quantification of multipartite entanglement is simplified for special classes of states possessing some symmetry such as
permutation invariance  \cite{Aulbach2010}.  These states can be created and characterized in multiqubit quantum experiments
with efficient tomographic protocols \cite{Toth2010}.

However, entanglement typically results to be very fragile with respect to the coupling with external degrees of freedom
\cite{BookBreuer, Yu04, BellomoPRL07}.
 A branch of the entanglement theory  is dedicated to finding
methods to contrast the  environment-induced degradation of entanglement \cite{Carvalho11,Gisin01,Lidar98,BellomoPRL07,Kim12,Viola99, Cirac09,Shapiro11,Dolde14}.
In particular, reservoir engineering methods have pointed out the possibility to change the perspective from reducing the coupling with the
environment to modifying the environmental properties in order to manipulate the system of interest thanks to its proper dissipative dynamics
 \cite{Plenio02,Hartmann06,Cirac09,Krauter11,Lucas2013}.

Recently, it has been studied the dynamics of few atoms driven by reservoirs at different temperatures, both in the case of ideal independent
 black-body thermal reservoirs  \cite{Quiroga07,Camalet2011, Wu11,Znidaric12,Camalet2013, Linden11} and in more realistic OTE configurations
 keeping into account  the scattering properties of macroscopic bodies held at different temperatures around one
  \cite{BellomoEPL2012, BellomoPRA2013} or two atoms \cite{BellomoEPL2013, BellomoNJP2013, Leggio15}. Similar approaches  have been
   also considered in the context of heat transfer \cite{MesAntEPL11,MesAntPRA11},  and Casimir-Lifshits forces
   \cite{AntezzaPRL05, AntezzaJPA06,ObrechtPRL07,AntezzaPRA08,Bimonte09,KardarPRL2011, MesAntEPL11,MesAntPRA11}.

In this paper we study if and how the steady state of more than two atoms can be controlled thanks to the dissipative dynamics
driven by a common OTE  field.  The attention will be mainly focused on the creation and control of quantum correlations
among the atoms due to the interaction between the atomic dipoles induced by the fluctuations of the common field.
We will study the dependence of this control on the spatial disposition of the atoms with respect to the bodies composing
their environment, on the internal properties of the atoms and of the bodies, such as the atomic frequencies and
the geometry and the dielectric properties of the bodies, not to mention the involved temperatures.
Permutationally invariant atomic states will play a fundamental role in this analysis.

\section{Model}

\begin{figure}[t!]
\begin{center} \includegraphics[width=0.425
\textwidth]{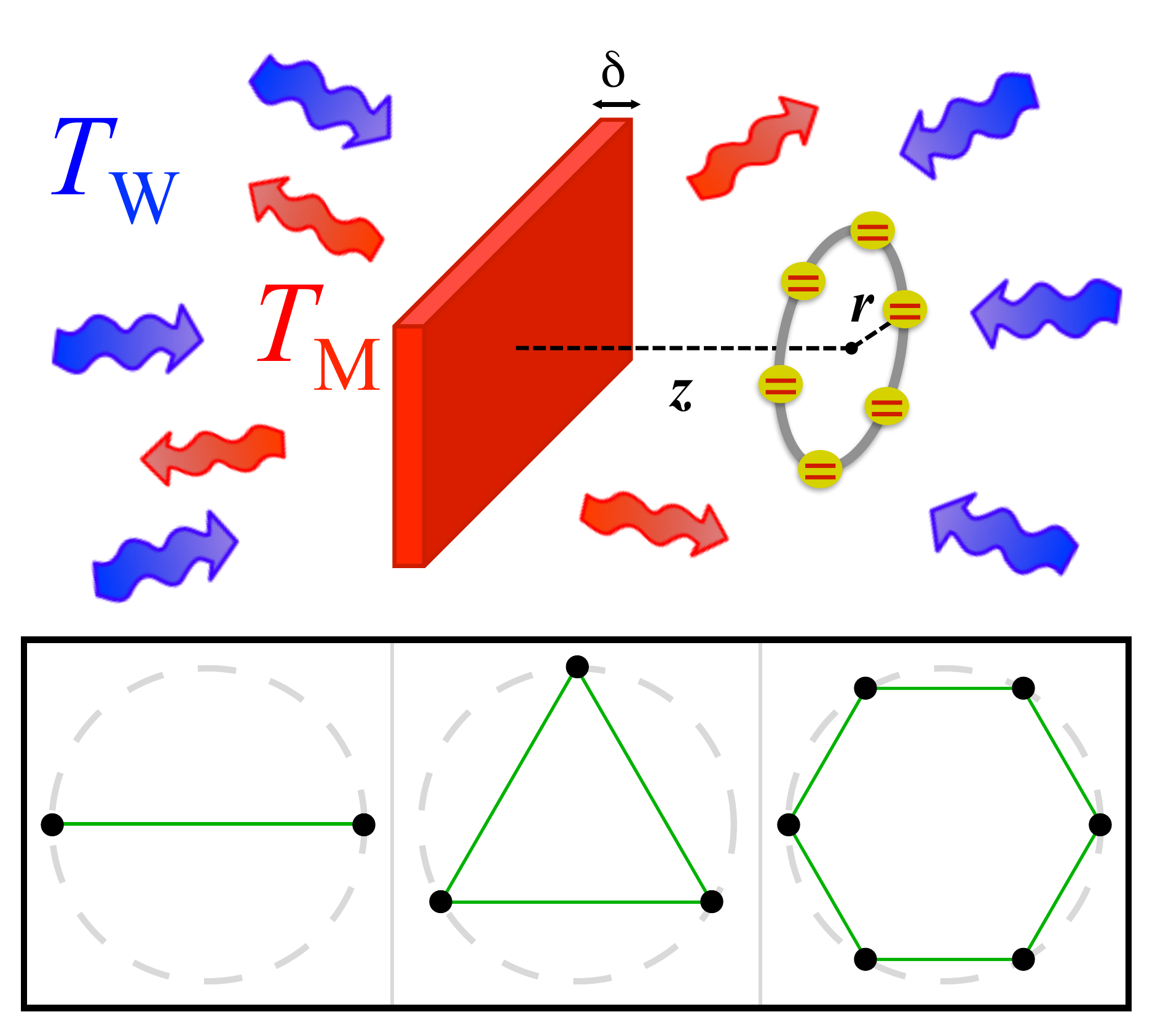} \end{center}
\caption{\label{fig1}\footnotesize (color online). The $N$ qubits are placed in a same plane x-y at a  distance $z$ from a dielectric slab of thickness $\delta$ whose temperature $T_M$ is kept constant and different form that of the surrounding blackbody radiation, $T_W$, which is also constant. At the bottom of the figure we show the  case when the qubits occupy the vertices of regular polygons inscribed in a circle of radius $r$,  for  $N=2, 3, 6$.}
\end{figure}

We want to study the dynamics of an open system $S$ made by an ensemble of $N$ two-level emitters (qubits) interacting with a common environment which is a stationary electromagnetic field out of thermal equilibrium (see Fig. \ref{fig1}). This OTE field is the result of the thermal radiation emitted by a macroscopic slab $M$ of thickness $\delta$, which is close to the qubits,  and by some far walls $W$ (not shown in the figure). The temperatures of the slab,  $T_\mathrm{M}$, and of the walls,  $T_\mathrm{W}$, are kept constant and in general different between them. The radiation emitted by the far walls can be safely treated at the qubits position as blackbody radiation in the absence of the slab. However, this radiation is eventually reflected and transmitted by the slab. The local field felt by each qubit is then the result of four contributions leading to a strong dependence from all the physical parameters involved. These parameters are  the common frequency  of the qubits and their positions with respect to the slab, the geometrical (thickness $\delta$) and dielectric (its resonances) properties of the slab $M$ and the two temperatures $T_\mathrm{M}$ and $T_\mathrm{W}$.

The total Hamiltonian is $H_T=H_S+H_E+H_I$, where $H_S$ and $H_E$ are the free Hamiltonians of the system $S$ and of the environment $E$ and $H_I$ represents their interaction.
 $S$ is composed by $N$ qubits having two internal levels $ \ket{g}_i$ and  $\ \ket{e}_i$ with the same transition frequency $\omega_0$. The free Hamiltonian of the qubits is $H_S=\sum_i \hbar \omega_0 \sigma_i^+ \sigma_i^-$ being $\sigma_i^+$ ($\sigma_i^-$) the raising (lowering) operator of the $i_{th}$ qubit. The interaction between $S$ and  $E$ is given in the multipolar coupling and in dipole approximation by
$H_I=$ $-\sum_i \mathbf{D}_i\cdot\mathbf{E}(\mathbf{R}_i)$ \cite{CohenTannoudji97},
where $\mathbf{D}_i$ is the electric-dipole operator of the $i_{th}$ qubit ($\mathbf{d}^i={}_i\bra{g}\mathbf{D}_i\ket{e}_i$ being its transition matrix element) and $\mathbf{E}(\mathbf{R}_i) $ is the electric field at its position $\mathbf{R}_i$.

Under Born, Markovian and rotating-wave approximations, a microscopic derivation for the master equation describing the  dynamics of the $N$ qubits gives \cite{BookBreuer, BellomoNJP2013}
\begin{equation}\label{master equation}\begin{split} \frac{d}{d t}\rho =& - \frac{i}{\hbar} [H_S +  \delta_S,\rho ]
-i\sum_{i \neq j}\Lambda_{i j}[\sigma_{i}^{+}\sigma_{j}^{-},\rho ]  \\  &
+\sum_{i, j}\Gamma_{ij}^+\Big(\sigma_{j}^{-}\rho \sigma_{i}^{+} -\frac{1}{2}\{\sigma^{+}_{i}\sigma^{-}_{j},\rho \}\Big)  \\  & + \sum_{i, j}\Gamma_{ij }^-\Big(\sigma^{+}_{j}\rho \sigma^{-}_{i}-\frac{1}{2}\{\sigma^{-}_{i} \sigma^{+}_{j},\rho \}\Big),\end{split}\end{equation}
where $
\delta_S=\sum_{i}\hbar  \bigl[S_{i}^+ -S_i^-\bigr]  \sigma_i^+ \sigma_i^-$. The functions $S_{i}^{\pm}$, $\Lambda_{ij}^{\pm}$ and $\Gamma_{ij}^{\pm}$ depend on qubits dipoles and on frequency (their explicit dependence on $\omega$ is omitted), they are evaluated at $\omega_0$ and are given by  \cite{BellomoEPL2013, BellomoNJP2013}
\begin{equation}\label{me parameters N}\begin{split}
 S_{i}^+&=\sum_{l,l'} s_{ll'}^{ii}(\omega)[\textbf{d}^i]^*_{l} [\textbf{d}^{i}]_{l'},\;
S_{i}^-=\sum_{l,l'}s_{ll'}^{ii}(-\omega)[\textbf{d}^i]_{l}[\textbf{d}^i]^*_{l'},
\\
 \Lambda_{ij}&= \sum_{l,l'}  [\mathbf{d}^i]_l^* [\mathbf{d}^{j}]_{l'}
 \bigl[s_{ll'}^{i j}(\omega)+s_{l'l}^{j i}(-\omega)\bigr],
\\
 \Gamma_{ij}^+&=\sum_{l,l'}\gamma_{ll'}^{ij}(\omega)[\textbf{d}^i]^*_{l} [\textbf{d}^{j}]_{l'},\;
\Gamma_{ij}^-=\sum_{l,l'}\gamma_{ll'}^{ij}(-\omega)[\textbf{d}^i]_{l}[\textbf{d}^{j}]^*_{l'},\end{split}\end{equation}
where $l$ and $l'$ $\in  \{x,y,z\}$ and the functions  $s_{ll'}^{ij}(\pm \omega)$ and  $\gamma_{ll'}^{ij}(\pm \omega)$ are connected  to the field correlation functions through the following relations
\begin{equation}\label{Xi function}\begin{split}
\gamma_{ll'}^{ij}(\omega) &=\Xi_{ll'}^{ij}(\omega) +\Xi_{l' l }^{j i \,*}(\omega)  , \; s_{ll'}^{ij}(\omega) =  \frac{\Xi_{ll'}^{ij}(\omega) -\Xi_{l' l }^{ji \,*}(\omega)}{2 i}, \\
\Xi_{ll'}^{ij}(\omega)&= \frac{1}{\hbar^2}\int_0^\infty  \!\!ds\,e^{i\omega s}\langle E_l(\mathbf{R}_i,s)E_{l'}(\mathbf{R}_{j},0)\rangle .\end{split}\end{equation}

\subsection{Master equation parameters out of thermal equilibrium}\label{par:ME parameters}

The functions $S_{i}^{\pm}$, $\Lambda_{ij}^{\pm}$ and $\Gamma_{ij}^{\pm}$ appearing in the master equation \eqref{master equation} are then connected to the field correlation functions, $\langle E_l(\mathbf{R}_i,s)E_{l'}(\mathbf{R}_{j},0)\rangle$, which  depend on all the physical parameters such as $\delta$, the qubits positions and the temperatures $T_\mathrm{M}$ and $T_\mathrm{W}$. The explicit form of these functions in the absence of thermal equilibrium has been derived  in \cite{BellomoEPL2013, BellomoNJP2013}. In particular, the transition rates $\Gamma_{ij }^+$ and $\Gamma_{ij }^-$ are given by:
\begin{equation}\begin{split}
\label{gamma functions finale}
\Gamma_{ij }^+=& \sqrt{\Gamma_0^i(\omega)\Gamma_0^{j}(\omega)} \Big\{[1+n(\omega,T_\mathrm{W})]\alpha_\mathrm{W}^{ij}(\omega)
\\& + [1+n(\omega,T_\mathrm{M})]\alpha_\mathrm{M}^{ij}(\omega)\Big\}\\
\Gamma_{ij}^-= & \sqrt{\Gamma_0^i(\omega)\Gamma_0^{j}(\omega) } \Big\{n(\omega,T_\mathrm{W})  \alpha_\mathrm{W}^{ij}(\omega)^*
\\&+n(\omega,T_\mathrm{M})]\alpha_\mathrm{M}^{ij}(\omega)^*\Big\}
 ,\end{split}\end{equation}
where
$
\alpha_\mathrm{W}^{ij}(\omega) =\sum_{l,l'}[\tilde{\textbf{d}}^i]^*_{l} [\tilde{\textbf{d}}^{j}]_{l'} [\alpha_\mathrm{W}^{ij}(\omega)]_{ll'} $, $
\alpha_\mathrm{M}^{ij}(\omega) =\sum_{l,l'}[\tilde{\textbf{d}}^i]^*_{l} [\tilde{\textbf{d}}^{j}]_{l'} [\alpha_\mathrm{M}^{ij}(\omega)]_{ll'}$,
being $[\tilde{\textbf{d}}^i]_{l}=[\textbf{d}^i]_{l}/|\textbf{d}^i|$,  and $\Gamma_0^i(\omega)=|\textbf{d}^i|^2\omega^3/3  \hbar  \pi\epsilon_0 c^3 $ is the vacuum spontaneous-emission rate of the $i_{th}$ qubit. The functions $[\alpha_\mathrm{M}^{ij}(\omega)]_{ll'} $ and $[\alpha_\mathrm{W}^{ij'}(\omega)]_{ii'} $ do not depend on temperatures, but they depend on all the other system parameters. In the general case in which the body close to the qubits has an arbitrary shape, by indicating the position of the $i_{th}$ qubit as $\mathbf{R}_i=(\mathbf{r}_i, z_i)$, the $\alpha$ functions can be cast under the form
\begin{equation}\label{alphaWM}\begin{split}
&[\alpha_\mathrm{W}^{ij}(\omega)]_{ll'}=\frac{3\pi c}{2 \omega}\sum_{p,p'}\int\frac{d^2\mathbf{k}}{(2\pi)^2}\int\frac{d^2\mathbf{k}'}{(2\pi)^2}e^{i(\mathbf{k}\cdot\mathbf{r}_i-\mathbf{k}'\cdot\mathbf{r}_{j})}\\ &
\, \times \bra{p,\mathbf{k}}\Bigl\{e^{i(k_z z_i-k_z^{'*}z_{j})}[\hat{\bbm[\epsilon]}_p^+(\mathbf{k},\omega)]_l[\hat{\bbm[\epsilon]}_{p'}^{+}(\mathbf{k}',\omega)]_{l'}^*\nonumber
\end{split}\end{equation}

\begin{equation}\begin{split}
&\,\times \Bigl(\mathcal{T}\mathcal{P}_{-1}^{\text{(pw)}}\mathcal{T}^{\dag}+\mathcal{R}\mathcal{P}_{-1}^{\text{(pw)}}\mathcal{R}^{\dag}\Bigr)+e^{i(k_z z_i+k_z^{'*}z_{j})} \\&
\, \times [\hat{\bbm[\epsilon]}_p^+(\mathbf{k},\omega)]_l[\hat{\bbm[\epsilon]}_{p'}^{-}(\mathbf{k}',\omega)]_{l'}^*\mathcal{R}\mathcal{P}_{-1}^{\text{(pw)}}+e^{-i(k_z z_i+k_z^{'*}z_{j})}\\
&  \, \times[\hat{\bbm[\epsilon]}_p^-(\mathbf{k},\omega)]_l[\hat{\bbm[\epsilon]}_{p'}^{+}(\mathbf{k}',\omega)]_{l'}^*\mathcal{P}_{-1}^{\text{(pw)}}\mathcal{R}^{\dag}+
e^{-i(k_z z_{i}-k_z^{'*}z_{j})} \\& \, \times [\hat{\bbm[\epsilon]}_p^-(\mathbf{k},\omega)]_l[\hat{\bbm[\epsilon]}_{p'}^{-}(\mathbf{k}',\omega)]_{l'}^*\mathcal{P}_{-1}^{\text{(pw)}}\Big\}\ket{p',\mathbf{k}'} ,
\\
&[\alpha_\mathrm{M}^{ij}(\omega)]_{ll'}=\frac{3\pi c}{2 \omega}\sum_{p,p'}\int\frac{d^2\mathbf{k}}{(2\pi)^2}\int\frac{d^2\mathbf{k}'}{(2\pi)^2}e^{i(\mathbf{k}\cdot\mathbf{r}_i-\mathbf{k}'\cdot\mathbf{r}_{j})}  \bra{p,\mathbf{k}}
\\ &  \, \Bigl\{e^{i(k_z z_i-k_z^{'*}z_{j})}[\hat{\bbm[\epsilon]}_p^+(\mathbf{k},\omega)]_l[\hat{\bbm[\epsilon]}_{p'}^{+}(\mathbf{k}',\omega)]_{l'}^*\Bigl[\Bigl(\mathcal{P}_{-1}^\text{(pw)} +\mathcal{R}\mathcal{P}_{-1}^\text{(ew)}
\\
&\, -\mathcal{P}_{-1}^\text{(ew)}\mathcal{R}^{\dag} -\mathcal{R}\mathcal{P}_{-1}^\text{(pw)}\mathcal{R}^{\dag}  -\mathcal{T}\mathcal{P}_{-1}^\text{(pw)}\mathcal{T}^{\dag}\Bigr)\,\Big\}\ket{p',\mathbf{k}'},\end{split}\end{equation}
where the operators $\mathcal{R}$ and $\mathcal{T}$ are the standard reflection and transmission scattering operators associated to the right side of the body, whose explicit definition can be found, for example, in \cite{MesAntPRA11}. They connect any outgoing (reflected or transmitted) mode of the field to the entire set of incoming modes.

In Eq. \eqref{alphaWM}, each mode of the field is identified by the frequency $\omega$, the transverse wave vector $\mathbf{k}=(k_x,k_y)$, the polarization index $p$ (taking the values $p=1,2$ corresponding to TE and TM polarizations respectively), and the direction of propagation $\phi=\pm1$ (shorthand notation $\phi=\pm$) along the $z$ axis. The total wavevector takes then the form $\mathbf{K}^\phi=(\mathbf{k},\phi k_z)$, where the $z$ component of the wavevector $k_z$ is a dependent variable given by $k_z=\sqrt{\frac{\omega^2}{c^2}-k^2}$, where $k=|\mathbf{k}|$.
The polarization vectors appearing in Eq. \eqref{alphaWM}  are defined in a standard way by
$
\hat{\bbm[\epsilon]}^\phi_\TE(\mathbf{k},\omega)=\hat{\mathbf{z}}\times\hat{\mathbf{k}}=(-k_y\hat{\mathbf{x}}+k_x\hat{\mathbf{y}})/k,
\hat{\bbm[\epsilon]}^\phi_\TM(\mathbf{k},\omega)=c\, \hat{\bbm[\epsilon]}^\phi_\TE(\mathbf{k},\omega)\times\mathbf{K}^{\phi}/\omega=c\, (-k\hat{\mathbf{z}}+\phi k_z\hat{\mathbf{k}})/\omega,
$
where $\hat{\mathbf{x}}$, $\hat{\mathbf{y}}$ and $\hat{\mathbf{z}}$ are the unit vectors along the three axes and $\hat{\mathbf{k}}=\mathbf{k}/k$. In Eq. \eqref{alphaWM} we have also used
$ \bra{p,\mathbf{k}}\mathcal{P}_n^\text{(pw/ew)}\ket{p',\mathbf{k}'}=k_z^n\bra{p,\mathbf{k}}\Pi^\text{(pw/ew)}\ket{p',\mathbf{k}'}$,
$\Pi^\text{(pw)}$ and $\Pi^\text{(ew)}$ being the projectors on the propagative ($ck<\omega$, corresponding to a real $k_z$) and evanescent ($ck>\omega$, corresponding to a purely imaginary $k_z$) sectors respectively.

Expressions more explicit for the $\alpha$ functions appearing in Eq. \eqref{alphaWM} have been derived in \cite{BellomoNJP2013} (see Sec. IV therein) by using the explicit form for the reflection and transmission operators in the case treated in this paper, that is when the body close to the qubits is a slab. In particular, because of the translational invariance of a planar slab with respect to the x-y plane, $\mathcal{R}$ and $\mathcal{T}$, are diagonal and given by
\begin{equation}\label{RT1slab}\begin{split} \bra{p,\mathbf{k}}\mathcal{R}\ket{p',\mathbf{k}'}&=(2\pi)^2\delta(\mathbf{k}-\mathbf{k}')\delta_{pp'}\rho_{p}(\mathbf{k},\omega),\\
\bra{p,\mathbf{k}}\mathcal{T}\ket{p',\mathbf{k}'}&=(2\pi)^2\delta(\mathbf{k}-\mathbf{k}')\delta_{pp'}\tau_{p}(\mathbf{k},\omega),
\end{split}\end{equation}
where the Fresnel reflection and transmission coefficients modified by the finite thickness $\delta$ have the form:
\begin{equation}\begin{split} \rho_{p}(\mathbf{k},\omega)&=r_{p}(\mathbf{k},\omega)\frac{1-e^{2ik_{zm}\delta}}{1-r_{p}^2(\mathbf{k},\omega)e^{2ik_{zm}\delta}},\\
\tau_{p}(\mathbf{k},\omega)&=\frac{t_{p}(\mathbf{k},\omega)\bar{t}_{p}(\mathbf{k},\omega)e^{i(k_{zm}-k_z)\delta}}{1-r_{p}^2(\mathbf{k},\omega)e^{2ik_{zm}\delta}},\\
\end{split} \end{equation}
where $k_{zm}$ is  the $z$ component of the $\mathbf{K}$ vector inside the medium,
$k_{zm}=\sqrt{\varepsilon(\omega)\frac{\omega^2}{c^2}-\mathbf{k}^2}$,
$\varepsilon(\omega)$ being the dielectric permittivity of the  slab. In the above equations we have also introduced the ordinary vacuum-medium Fresnel reflection coefficients
\begin{equation}r_{\mathrm{TE}}=\frac{k_z-k_{zm}}{k_z+k_{zm}},\qquad r_{\mathrm{TM}}=\frac{\varepsilon(\omega)k_z-k_{zm}}{\varepsilon(\omega)k_z+k_{zm}},\end{equation}
as well as both the vacuum-medium (noted with $t$) and medium-vacuum (noted with $\bar{t}$) transmission coefficients
\begin{equation}\begin{split} t_{\mathrm{TE}}&=\frac{2k_z}{k_z+k_{zm}},\qquad\hspace{.3cm}t_{\mathrm{TM}}=\frac{2\sqrt{\varepsilon(\omega)}k_{z}}{\varepsilon(\omega)k_z+k_{zm}},\\
\bar{t}_{\mathrm{TE}}&=\frac{2k_{zm}}{k_z+k_{zm}},\qquad\bar{t}_{\mathrm{TM}}=\frac{2\sqrt{\varepsilon(\omega)}k_{zm}}{\varepsilon(\omega)k_z+k_{zm}}.\end{split} \end{equation}

With regards to the $\Lambda_{ij}$ function appearing in Eq. \eqref{master equation}, it can be also expressed in terms of $\alpha$ functions as
\begin{equation}\label{lambda2}\begin{split}
&\Lambda_{ij}= \frac{\sqrt{\Gamma_0^i(\omega)\Gamma_0^{j}(\omega) }}{\omega^3}
\\ & \times  \mathcal{P}\int_{-\infty}^{+\infty}\frac{\omega'^3 d\omega'}{2\pi}\frac{\alpha_\mathrm{W}^{ij}(\omega') +\alpha_\mathrm{M}^{ij}(\omega') }{\omega-\omega'}.
 \end{split}
 \end{equation}

Concerning the operator $\delta_S$ in Eq.\eqref{master equation}, it is responsible for the shifts of energy levels such that the renormalized transition frequency of the $i_{th}$ qubit is equal to  $\omega_i=\omega_0+S_{i}^+-S_{i}^-$.  It is possible to show that only the differences between the shifts of each qubit influence the steady state. In the case when the body is a slab, their influence disappears when identical qubits are placed at the same distance from the slab, being all the shifts equal between them. In the following we will always be in this case and we will name with $z$  the common distance from the slab and with  $\tilde{\omega}_0$ the common renormalized transition frequency.

\subsection{Choice of the basis}\label{par:basis}

To give a simple interpretation of the qubits dynamics we consider a  basis obtained by extending  the approach used in \cite{Freedhoff04} at zero temperature. We first recast  master equation \eqref{master equation} under the form
\begin{equation}\label{master equation 2} \begin{split} \frac{d}{d t}\rho =&\frac{ i}{\hbar} \left( \rho \, H_{\mathrm{eff}} -H_{\mathrm{eff}} ^\dag \,\rho\right)
+\sum_{i, j}\Big[ \Gamma_{ij}^+ \sigma_{j}^{-}\rho \sigma_{i}^{+} + \Gamma_{ij }^-\sigma^{+}_{j}\rho \sigma^{-}_{i}\Big],\end{split}\end{equation}
where $H_{\mathrm{eff}} $ is a not hermitian effective Hamiltonian,
\begin{equation}\label{M} \begin{split} H_{\mathrm{eff}} &= \hbar
\sum_i \Big[ \Big( \omega_i+\frac{1}{2} \, i \,  \Gamma_{ii}^+\Big) \sigma_i^+ \sigma_i^- +\frac{1}{2} \, i \,  \Gamma_{ii}^- \sigma_i^- \sigma_i^+ \Big] \\  & +  \hbar
\sum_{i \neq j} \Big[ \Big( \Lambda_{ij}+\frac{1}{2} \, i \,  \Gamma_{ij}^+\Big) \sigma_i^+ \sigma_j^- +\frac{1}{2} \, i \,  \Gamma_{ij}^- \sigma_i^- \sigma_j^+ \Big],\end{split}\end{equation}
and where we used  $[\Lambda_{ij}]^*=\Lambda_{ji}$ and $[\Gamma_{ij}^{\pm}]^*=\Gamma_{ji}^{\pm}$.

The $H_S$ operator permits to partition the state vector space $W$ as $W= \otimes_{n=0}^N W^n$, where the $N+1$ sectors $W^n$  are spanned by the decoupled eigenstates of $H_S$ having eigenvalue $n \hbar \omega_0$ ($n$ qubits in the excited state and the $N-n$ remaining ones in the ground state) having a degeneracy (giving the dimensionality of $W^n$) equal to $d_n=N!/[(N-n)! n!]$. The interaction terms in the second line of Eq. \eqref{M} split this degeneracy. Under the action of $H_{\mathrm{eff}} $  the number of excitations is preserved because $H_{\mathrm{eff}} $  commutes with $n_{exc}=\sum_{i=1}^N \sigma_i^+ \sigma_i^-$, representing the number of excited qubits in the system ($[H_{\mathrm{eff}} ,n_{exc}]=0$).
For each number of excitations $n$, $W^n$ can thus be spanned  in terms of the collective eigenstates of $H_{\mathrm{eff}} $. These states can be found for each $n$ by solving the eigenvalue equation of the ($d_n$ x $d_n$)  matrix  representing $H_{\mathrm{eff}} $ in the space spanned by the corresponding $d_n$ decoupled eigenstates of $H_S$. Because this matrix is not hermitian its eigenvalues and eigenvectors are in general complex (right and left eigenvectors in general do not coincide). For the extremal cases $n=0$ and $n=N$ the dimensionality is $d_0=d_N=1$ and the eigenstates  of $H_{\mathrm{eff}} $  correspond to the case when all the qubits are either in the ground or in the excited state.

In Appendix \ref{appendix} we show that the projection of master equation \eqref{master equation} on the basis of eigenstates of $H_{\mathrm{eff}} $
leads to the simplification that each density matrix element in $W^n$ is connected to all the populations and coherences
of $W^{n+1}$ and $W^{n-1}$ but not to the other elements of  $W^n$. One can further simplify the problem by using a secular
 approximation \cite{CohenTannoudji97, Freedhoff04} to decouple populations and coherences between not degenerates eigenstates.
 Following \cite{Freedhoff04}, the real part of the complex eigenvalue of $H_{\mathrm{eff}} /\hbar$  gives the frequency shift of the collective state while twice the imaginary part gives the total decay constant of the state (or inverse lifetime), that is the sum of the individual decay constants towards all the states in the energy manifolds below and above.

In the following, we investigate  in the coupled basis  the steady  correlations built up between the $N$ qubits during the dissipative dynamics. To this purpose we numerically integrate the master equation in the decoupled basis, where the only block of coupled equations involving density matrix elements which can be steadily different from zero concerns all the populations and all the coherences between states having the same number of excitations.

\section{Correlations and their quantification}

In the case of a dissipative dynamics one unavoidably deals with mixed states which still can contain some
entanglement.   Its quantification is considered to be completely solved only in the simplest case of a two-qubit system. For higher dimensional non-pure states, in general, the problem
of characterization of the set of separable mixed states appears to be extremely complex \cite{Horodecki09}.

As main quantifier of entanglement we will make use  of negativity. This quantity is strictly related to the Peres-Horodecki criterion \cite{Peres96}, which gives a necessary condition for an arbitrary bipartite state $\rho_{AB}$ to be separable. This separability condition is also sufficient in the $2\times 2$ and $2\times 3$  dimensional cases and  is given by the positivity  of the partial transpose, which is obtained from any given bipartite quantum state by transposing
the variables of only one of the two subsystems.

The negativity $\mathcal{N}$  associated to an arbitrary  state $\rho_{AB}$ is defined as
\begin{equation}\label{negativity}
\mathcal{N}(\rho_{AB})=-2 \sum_i \sigma_i (\rho^{TA}),
\end{equation}
where ${\sigma_i (\rho^{TA})}$ are the negative eigenvalues of the partial transpose $\rho^{TA}$ of  $\rho_{AB}$ respect to the subsystem A, defined as $\bra{i_A,j_B} \rho^{TA}\ket{k_A,l_B}=\bra{k_A,j_B}
\rho_{AB} \ket{i_A,l_B}$ (note that the definition of $\mathcal{N}$ is independent of the part that is transposed). We use here as negativity twice the original definition, so that it ranges from 0 for a separable state to 1 for a maximally entangled state. Negativity is an entanglement monotone (including convexity) and thus can be considered a useful measure of entanglement  \cite{Vidal2002}.

In this paper we consider different forms of entanglement associated to a state of  $N$ qubits. First, we quantify the entanglement between two qubits $i$ and $j$  by tracing out the remaining qubits and by using the two-qubit negativity  $\mathcal{N}_{i-j}$. We will then compare its value with another entanglement quantifier, the concurrence  $C_{i-j}$ \cite{Wootters98}. Concurrence is also an entanglement monotone, with a  direct physical meaning due to its connection with  the entanglement of formation \cite{Horodecki09}. Second, we  quantify entanglement between groups of qubits by analyzing the negativity associated to all the possible bipartitions $(\dots i j k \dots / \dots lmn\dots)$ obtained without tracing out any qubits ($\mathcal{N}_{\dots i j k\dots /\dots lmn\dots}$).

Finally, in the case of three qubits ($N=3$) we make use of the tripartite negativity defined as \cite{Sabin2008}
\begin{equation}\label{tripartitenegativity}
\mathcal{N}_{123}=\left(\mathcal{N}_{1/23}\mathcal{N}_{2/13}\mathcal{N}_{3/12}\right)^{\frac{1}{3}}.
\end{equation}
Indeed,  a three-qubit non-pure state $\rho$ may be fully separable ($\rho=\sum_i p_i\rho_i^A\otimes \rho_i^B\otimes \rho_i^C$), biseparable ($\rho=\sum_{iJ} p_{iJ} \rho_i^J\otimes \rho_i^{KL}$, where $J$ runs from $A$ to $C$ and $KL$ from $BC$ to $AB$, and at least one $\rho_i^{KL}$ is entangled), or fully inseparable (not fully separable nor biseparable). In particular, simply biseparable states have $p_{iJ}= 0$ only for a single value of $J$, corresponding to the case where one single qubit is separable from the other two, that are entangled. Differently, in the case of  generalized biseparable states there are non vanishing coefficients $ p_{iJ}$ for more than one $J$. Fully inseparable states have fully tripartite (genuine) entanglement which may be quantified by $\mathcal{N}_{123}$ of Eq. \eqref{tripartitenegativity}, as shown in \cite{Sabin2008}. However, it is known that $\mathcal{N}_{123}$ could be non-zero for generalized biseparable
states so that a finite value of the tripartite negativity for a given state, even if it assures that this state is not biseparable with respect to any bipartition, it does not exclude the possibility the state being a convex mixture of states that are biseparable with respect to different bipartitions.

\section{Entanglement out of thermal equilibrium}

Recently, we have shown that two qubits interacting with a common OTE field may thermalize into entangled states, differently from the case at thermal equilibrium where the entanglement induced by the common environment vanishes off asymptotically \cite{Benatti03}. In the following, we investigate the entanglement properties of $N$-qubits steady states for fixed $N=3, 6$ (chosen as illustrative cases) and their functional dependence on $N$. We compare the amount of entanglement generated with more than two qubits with respect to the case when only two qubits are present.

In all the following numerical analysis we choose the sapphire, characterized by a first resonance at $0.81\times 10^{14}$ rad/s, as material for the  slab close to the qubits. This allows us to obtain the necessary average number of thermal photons setting one of the two temperatures equal to 300 K. The optical data for the sapphire dielectric permittivity are taken from \cite{Palik1998}. All the qubits are considered identical with dipole moments oriented along the $z$ direction.

The results obtained in the following strongly depend on the choice of the two temperatures $T_\mathrm{W}$ and $T_\mathrm{M}$. While all the forms of steady entanglement disappear when $T_\mathrm{W}=T_\mathrm{M}$, we will choose out of equilibrium configurations maximizing the entanglement in the limit case of only two qubits.

\subsection{Three qubits}

\begin{figure}[t!]
\begin{center} \includegraphics[width=0.36\textwidth]{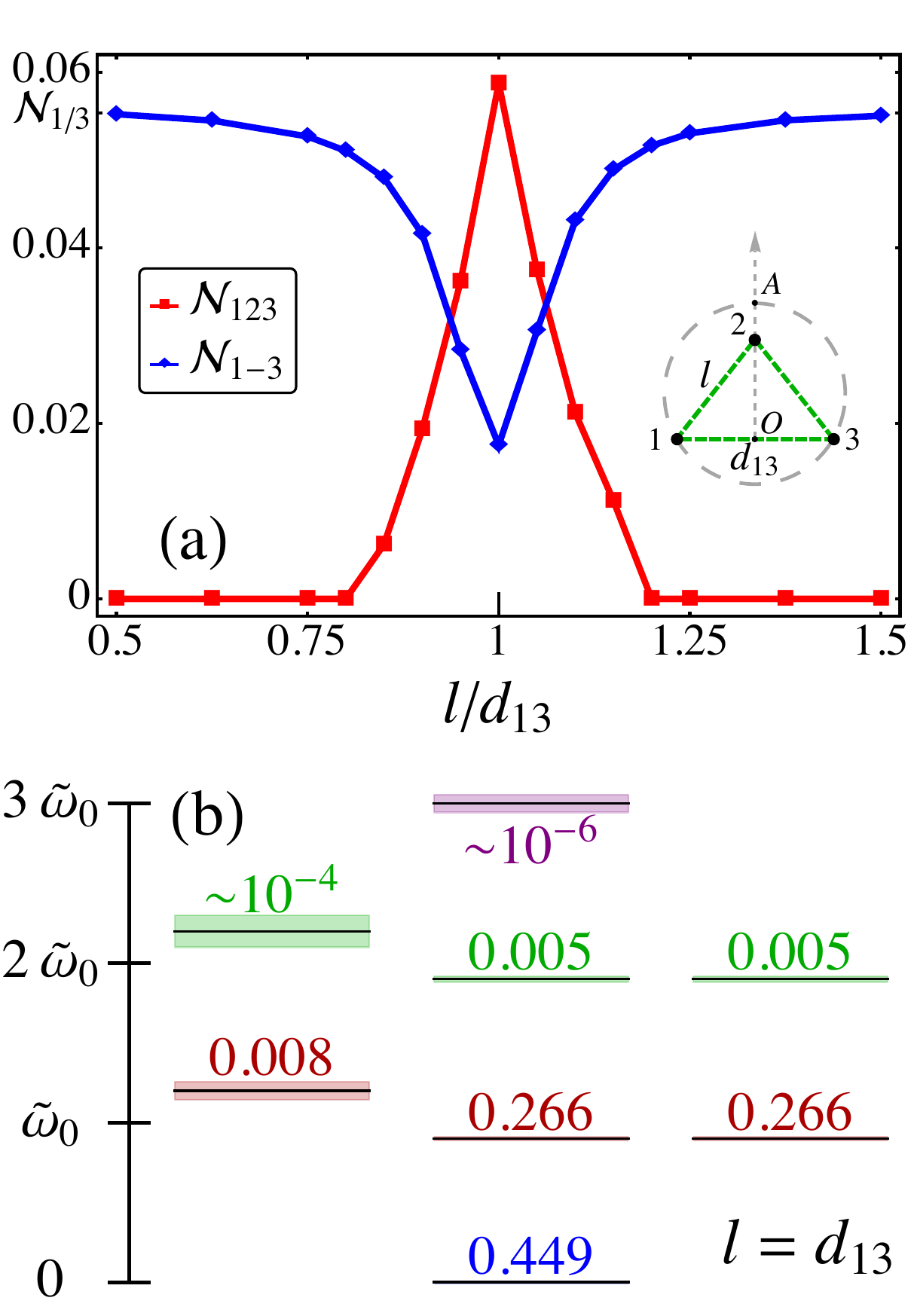} \end{center}
\caption{\label{fig2}\footnotesize (color online). The three qubits lie on a plane parallel to the slab of $\delta= 0.01\mu$m at a distance $z=8 \mu$m. The qubits frequency is fixed at $\omega_0=0.05\times 10^{14}$ rad$/$s, and the two temperatures at $T_\mathrm{W}=5$ K and  $T_\mathrm{M}=300$ K. Panel (a):  qubits 1 and 3 are kept fixed at a distance $d_{13}=2 \mu$m. Qubit 2 is moved along the line passing from $O$ and $A$. When qubit 2 is in $O$, the three qubits are aligned  and $l=1 \mu$m ($l/d_{13}=0.5$). When qubit 2 is in $A$,  the three qubits form an equilateral triangle of side  $l=2 \mu$m ($l/d_{13}=1$). The steady global negativity, $\mathcal{N}_{123}$, and the steady negativity between the external qubits 1 and 3 after tracing out qubit 2, $\mathcal{N}_{1-3}$, are plotted as a function of $l/d_{13}$. Panel (b): for the case $l=d_{13}$, we report populations and scaled eigenvalues of the collective states. The real part of the eigenvalues gives the position with respect to the various frequencies $n\,\tilde{\omega}_0$, while the imaginary part is connected to the lifetime of the different eigenstates, whose inverse is also depicted in scaled way.}
\end{figure}

We start our analysis by considering the case of three qubits ($N=3$), as shown in Fig.  \ref{fig2}. Our reference case is when qubit 2 is absent, which is characterized by a steady negativity and concurrence indicated by $\mathcal{N}_{1/3}$ and $C_{1/3}$.  The main aim of this section is to discuss the strong dependence of the steady entanglement production on the spatial disposition of the qubits.

We plot in panel (a) of Fig. \ref{fig2} the tripartite negativity $\mathcal{N}_{123}$ and the two-qubit negativity $\mathcal{N}_{1-3}$, obtained after tracing out the qubit 2, as a function of the ratio $l/d_{13}$, where $l$ is the distance between the second qubit and the external qubits 1 and 3, and $d_{13}$ is the distance between qubits 1 and 3. The three qubits are initially aligned  $(l/d_{13}=0.5)$ and then qubit 2 is moved along the vertical to the line.
The plot shows that tripartite genuine correlations measured by $\mathcal{N}_{123}$ rapidly switch on when the equilateral triangle configuration is approached ($l=d_{13}$). In this case, the symmetry in the steady state imposes that the bipartite negativities between qubit $i$ and the other two qubits $jk$  ($\mathcal{N}_{i/jk}$) are all equal for any choice $i$ and $jk$ so that each bipartite negativity $\mathcal{N}_{i/jk}$ coincides with the tripartite negativity $\mathcal{N}_{123}$  (see Eq. \eqref{tripartitenegativity}). When one moves from this configuration, correlations are mainly shared between external qubits, as  shown by the behavior of the two-qubit negativity  $\mathcal{N}_{1-3}$ which falls down for $l=d_{13}$, showing that steady states may present more tripartite entanglement than two-qubit one. In general, the presence of qubit 2 implies a reduction of the two-qubit negativity shared by the external qubits. If we only consider qubits 1 and 3, we obtain for the two-qubit negativity the value $\mathcal{N}_{1/3}\approx 0.055$, corresponding to a concurrence $C_{1/3}\approx 0.258$, comparable with the highest values found in \cite{BellomoEPL2013, BellomoNJP2013}.

Panel (b)  shows for the case $l=d_{13}$, i.e. when the three qubits form an equilateral triangle, the energy positions of the collective states for each sector  at fixed number of excitations, $W^{n}$, $n \in \{0,1,2,3\}$. The values of the steady populations in the collective basis are also reported (note that in this basis the steady state is diagonal and that the degenerate states in each sector are equally populated). Total decay constant of the collective states are also depicted, showing that typically the states with a larger lifetime are the ones steadily more populated. At thermal equilibrium,  the collective states of a given sector are always equally populated (qubits thermalise in their free hamiltonian eigenbasis), leading to steady thermal non-entangled states. Out of equilibrium, remarkably, collective states of the same $W^{n}$ can have different steady  populations  implying the emergence  of  steady quantum correlations.

Hence, the presence and the amount  of steady entanglement depends on two simultaneous mechanisms: on the one hand the absence of equilibrium allows to differently populate collective states  which, on the other hand, present an amount of entanglement strongly dependent on the spatial disposition
of the qubits. In particular, the regular polygon
configurations lead to strongly entangled collective states
offering the best scenario for entanglement production as
discussed below in the case of an equilateral triangle.

\subsubsection{Equilateral triangle\label{sec:triangle3q}}

We treat here the case when the three qubits form an equilateral triangle ($l=d_{13}$), by adding the condition that all the $\Gamma^{\pm}_{ij}$ are real (this occurs when the dipole moments of the qubits are real and have components either only along the z-axis or only along the plane x-y). This geometrical disposition implies that $\Gamma_{11}^{\pm}=\Gamma_{22}^{\pm}=\Gamma_{33}^{\pm}\equiv \Gamma^{\pm}$ and $\Gamma_{12(21)}^{\pm}=$ $\Gamma_{13(31)}^{\pm}=\Gamma_{23(32)}^{\pm}\equiv \Gamma_{q}^{\pm} $ and $\Lambda_{12(21)}=\Lambda_{13(31)}=\Lambda_{23(32)}\equiv \Lambda$.
As shown in panel (b),  the eigenstates  of $ H_{\mathrm{eff}} /\hbar$ mainly involved in the steady state are the ones  with zero or one excitations. These eigenstates, $\lambda$, and their eigenvalues,  $\Omega$, are

\begin{equation}\begin{split} \nonumber
&\ket{\lambda_1^{(0)}} = \ket{000},
\quad \ket{\lambda_1^{(1)}}  =    ( \ket{001}+\ket{010}+\ket{100})/\sqrt{3},\\
&\ket{\lambda_2^{(1)}}  = \frac{- \ket{001}+\ket{010}}{\sqrt{2}},   \quad
 \ket{\lambda_3^{(1)}}  =   \frac{- \ket{001}+\ket{100}}{\sqrt{2}},
 \end{split}\end{equation}

 \begin{equation}\label{M eigenstates 3 qubit triangle}\begin{split}
&\Omega_1^{(0)}= i\frac{3}{2} \Gamma^-, \, \Omega_1^{(1)}=  \tilde{\omega}_0 +2 \Lambda + i  \frac{\Gamma^++2( \Gamma^- +\Gamma^-_{q}+\Gamma^+_{q} )}{2}, \\
&  \Omega_2^{(1)}=  \tilde{\omega}_0 - \Lambda+ i \, \frac{2\Gamma^- + \Gamma^+-\Gamma^-_{q}-\Gamma^+_{q} }{2}, \quad \Omega_3^{(1)}= \Omega_2^{(1)}.
\end{split}\end{equation}

Eq. \eqref{M eigenstates 3 qubit triangle} shows that collective states may be strongly entangled  as a consequence of the permutational invariance of the qubits configuration ($\ket{\lambda_1^{(1)}}$ is for example a maximally entangled state). In Fig.  \ref{fig2}, for $l=d_{13}$, entanglement derives essentially by a classical mixing of the degenerate eigenstates $\ket{\lambda_2^{(1)}} $ and $\ket{\lambda_3^{(1)}}  $ (equal populations $\approx 0.266$) which have a smaller decay constant  than $\ket{\lambda_1^{(1)}}$, which is then much less populated ($\approx 0.008$).  We note that the amount of entanglement produced is strongly limited by the large population of the ground state $\ket{\lambda_1^{(0)}}$ ($\approx 0.449$).

It is easy to show that  when the three qubits are aligned ($l/d_{13}=0.5$), collective states in $W^{1}$ do not include the maximally entangled state $\ket{\lambda_1^{(1)}}$, and that the eigenstate more populated in this sector  has the form of a product state between the ground state of qubit 2 and a maximally entangled state between qubits 1 and 3, i.e. $(\ket{100}-\ket{001}/\sqrt{2})$. This is the reason why  $\mathcal{N}_{1-3}$ (keeping into account the presence of qubit 2 which is then traced out) is close to  $\mathcal{N}_{1/3}$ (obtained in absence of qubit 2).

We finally remark that, in Fig.\ref{fig2}, by keeping fixed $T_\mathrm{M}$ and by increasing $T_\mathrm{W}$ up to $\approx19 $K the tripartite negativity $\mathcal{N}_{123}$ goes to zero. The different forms of steady entanglement found in the next sections will share a similar dependence on $T_\mathrm{W}$.

\subsection{Six qubits}

\begin{figure}[t!]
\begin{center} \includegraphics[width=0.41\textwidth]{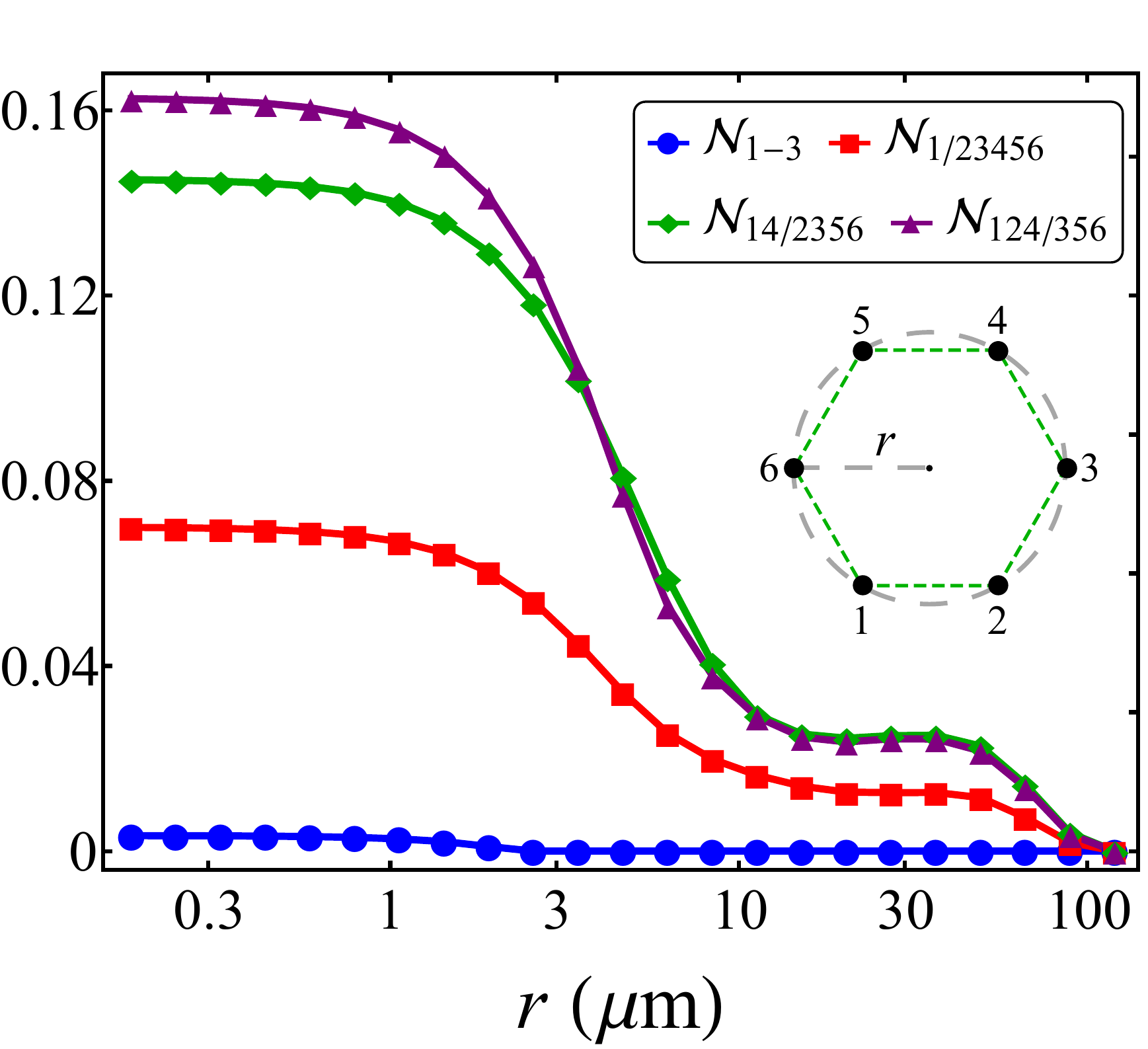} \end{center}
\caption{\label{fig3}\footnotesize (color online). The six qubits form a regular hexagon parallel to the slab and inscribed in a circle of radius $r$. The values  of $\delta, z, \omega_0, T_\mathrm{W}$ and  $T_\mathrm{M}$ are the same of Fig. \ref{fig2}. The maximal negativity among two-qubit negativities ($\mathcal{N}_{1-3}$) and among each type of bipartite negativities  ($\mathcal{N}_{1/23456}$, $\mathcal{N}_{14/2356}$ and $\mathcal{N}_{124/356}$) are plotted as a function of the circle radius $r$.}
\end{figure}

\begin{figure}[t!]
\begin{center} \includegraphics[width=0.41\textwidth]{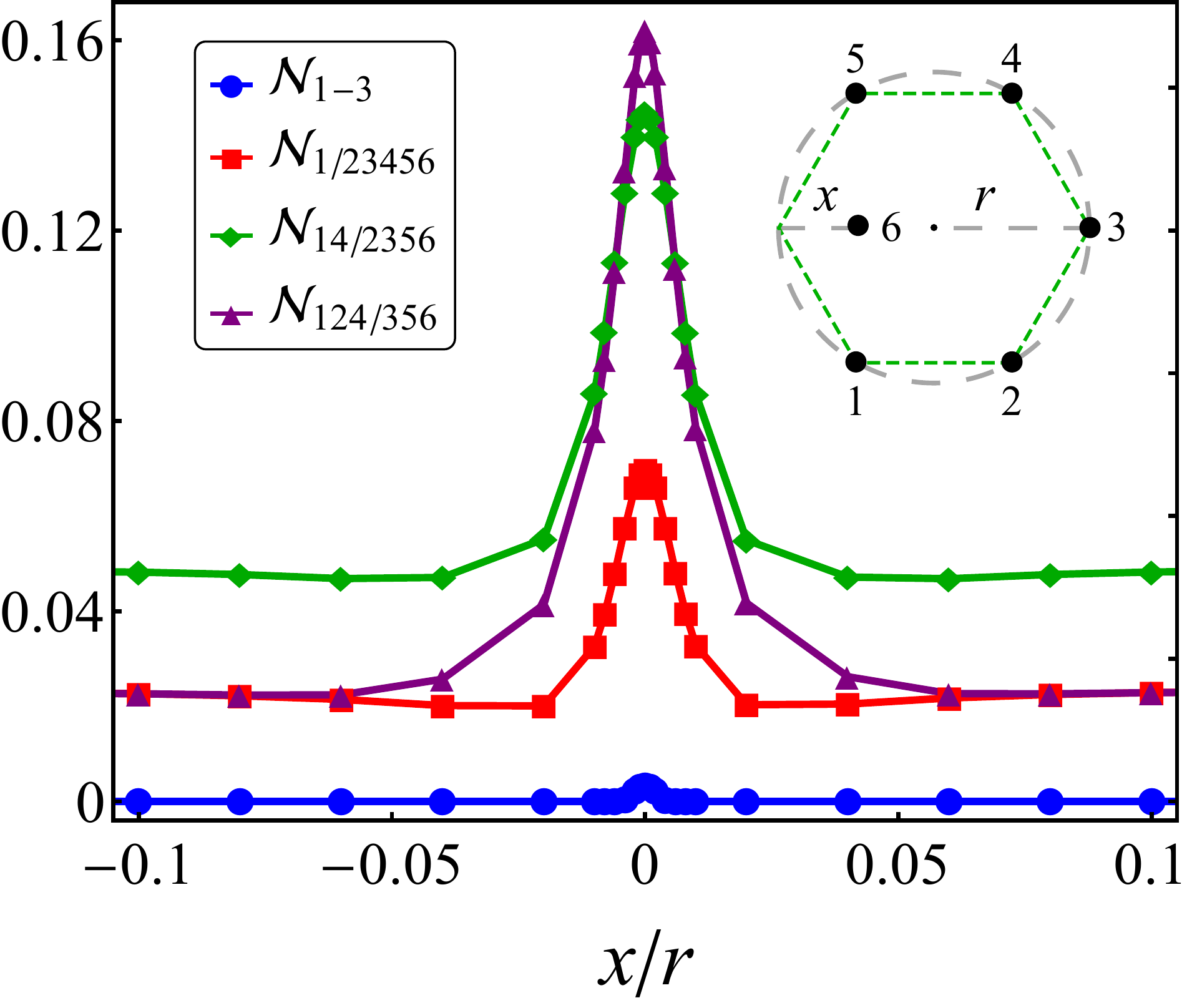} \end{center}
\caption{\label{fig4}\footnotesize (color online). The six qubits almost form a regular hexagon parallel to the slab and inscribed in a circle of radius $r\approx \,0.33 \mu$m. The values  of $\delta, z, \omega_0, T_\mathrm{W}$ and  $T_\mathrm{M}$ are the same of Fig. \ref{fig3}. The same negativities of Fig. \ref{fig3} ($\mathcal{N}_{1-3}$, $\mathcal{N}_{1/23456}$, $\mathcal{N}_{14/2356}$ and $\mathcal{N}_{124/356}$)  are plotted as a function of the ratio $x/r$. For $x=0$, a regular hesagon is formed and the corresponding maxima of negativities coincide with the values found in Fig. \ref{fig3} for $r\approx 0.33 \, \mu$m}.
\end{figure}

Here we consider the case of six qubits ($N=6$) forming a regular  hexagon parallel to the slab. The aim is to discuss the different forms of entanglement present in the steady state of the qubits. We focus on all the two-qubit negativities obtained tracing out four of the six qubits and the negativities of all the possible bipartitions obtained by dividing the six qubits in two groups made of, respectively, $n$ and  $6 - n$ qubits ($n\in\{1,2,3\}$). The symmetry in the total state makes many between these negativities equal. For example, there are only three not equivalent two-qubit negativities, i.e. $\mathcal{N}_{1-2}$, $\mathcal{N}_{1-3}$ and $\mathcal{N}_{1-4}$, being the choice of the first qubit free and since there are only three different qubit-qubit distances. Similar considerations hold for the bipartite negativities.

Among the two-qubit and the bipartite negativities for any $n$, we plot in Fig. \ref{fig3}  the maximal values as a function of the radius of the circle circumscribing the hexagon. The plot evidences an increasing of negativity for larger values of $n$  with values up to around 0.16, much larger than the ones obtained in the case of 3 qubits. A second important feature is a strong resistance of negativity with respect to the increase of distances between qubits. Negativity is still larger than zero up to radius of 100 $\mu$m. We also note the presence of a large plateau at large $r$ where negativity remains stable even if at small values. This plateau appears to be connected to the behaviour of the decay rates $\Gamma_{ij}^{-}$. For $r$ ranging between 15 and 35 $\mu$m, we numerically find that $\Gamma_{12}^{-}$, $\Gamma_{13}^{-}$ and $\Gamma_{14}^{-}$ are almost equal even if the distance between these couples of qubits varies. This differs from  what typically occurs for other values of $r$ and for $\Gamma_{ij}^{+}$. The origin of this behaviour of $\Gamma_{ij}^{-}$ can be found in the peculiar dependence on the distance between qubits of the integral describing the effects of the evanescing field emitted by the slab on the qubits dynamics \cite{BellomoNJP2013}.

In order to investigate how a deviation from the symmetric configuration of a regular hexagon affects the above results, we plot in Fig. \ref{fig4} the same negativities selected in Fig. 3 as a function of the ratio $x/r$, where $r$ is the radius of the circle depicted in the inset of this figure and $x$ is the distance of the sixth qubit from the position it should have to form a regular hexagon. This plot shows that negativities rapidly decrease when the symmetric configuration is lost, highlighting the fundamental role played by symmetry in the generation of multipartite steady entanglement out of thermal equilibrium. We remark that the entanglement reduction when one moves away from symmetric configurations, appears to be more pronounced of what already observed in Fig. \ref{fig2} in the case of three qubits.

\begin{figure}[t!]
\begin{center} \includegraphics[width=0.41\textwidth]{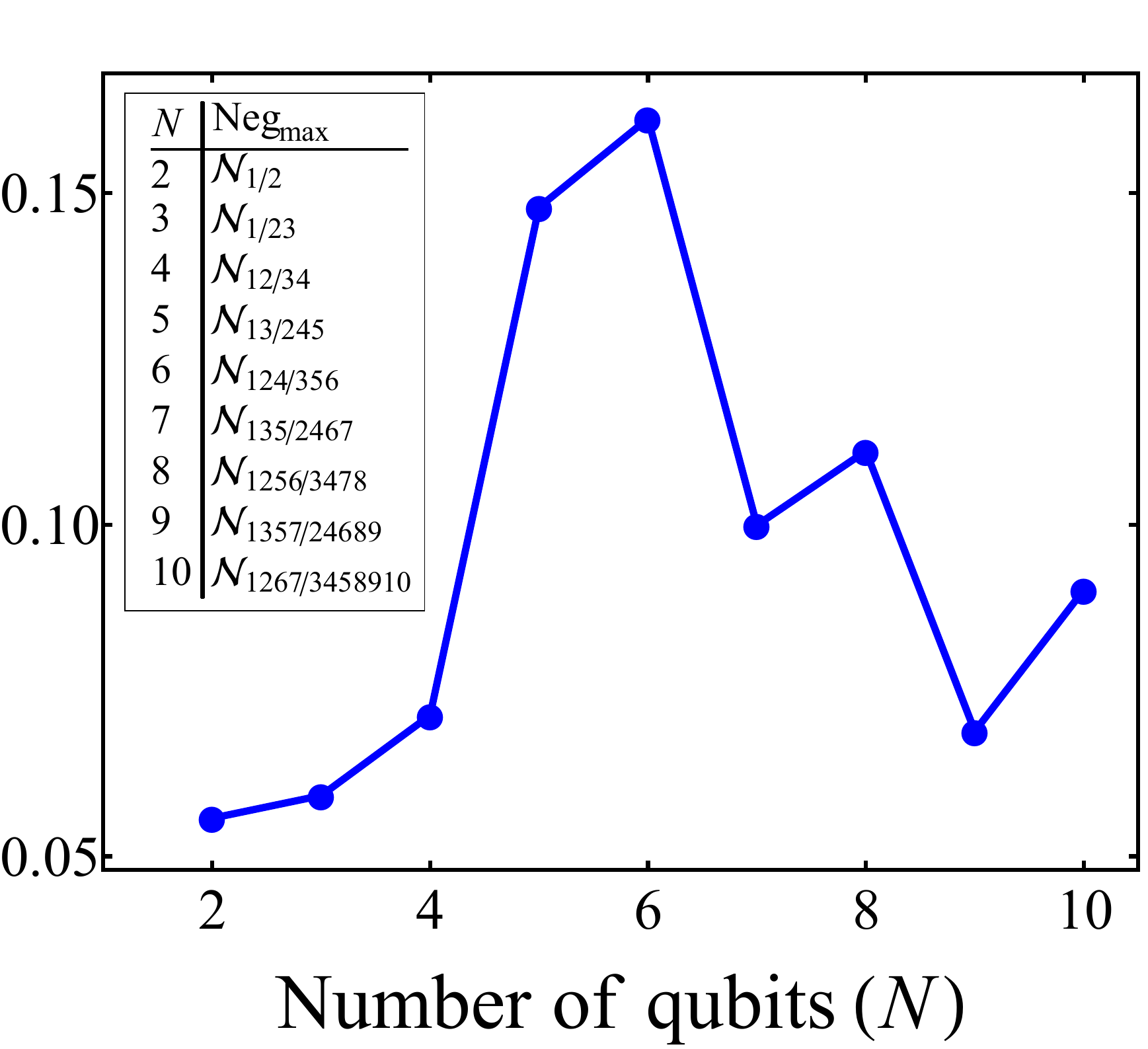} \end{center}
\caption{\label{fig5}\footnotesize (color online). For each $N$, the qubits form a regular polygon inscribed in a circle of radius 0.5 $\mu$m, parallel to the slab. The values  of $\delta, z, \omega_0, T_\mathrm{W}$ and  $T_\mathrm{M}$ are the same of Fig. \ref{fig2}. The maximal negativity is plotted as a function of the
number of qubits $N$, which varies between 2 and 10. For each $N$ the maximal negativity is found between all the two-qubit negativities and all the bipartite negativities.}
\end{figure}

\subsection{Arbitrary $N$}

Here we let the number of qubits $N$ vary, while being always placed at the vertices of regular polygons.
In Fig. \ref{fig5}, we plot as a function of the number of qubits,  the maximal negativity among all the possible two-qubit
negativities obtained by tracing out $N-2$ qubits and among all the possible bipartite negativities.
By increasing the number of qubits we find large values of negativities which seem to remain quite stable with a peak in the
case of the  hexagon ($N=6$). These values represent a lower threshold  to the maximal negativity obtainable for each
 configuration, being open the possibility to improve them by optimizing the various parameters involved in the system
 (we use here the values of the parameters optimizing in the three-qubit configuration the two-qubit negativity $\mathcal{N}_{1/3}$ (see fig. \ref{fig2})).

\section{Conclusions}

We studied the dynamics of an ensemble of qubits interacting with a common stationary field out of thermal
equilibrium resulting from  a rich yet simple configuration involving macroscopic bodies held at different
 temperature, which is within experimental reach.

We pointed out the possibility to generate steady entangled multipartite states giving a simple interpretation of the dynamics in a particular  collective basis. The absence of equilibrium permits to differently populate  collective states which at thermal equilibrium
would be steadily equally populated. This leads to the creation of different forms of steady  many-body
entanglement  strongly dependent on the characteristics of the collective states and on how they are populated. In the case of permutationally invariant qubits configurations, the collective states present the most of
  entanglement offering thus the ideal case to be exploited out of equilibrium.

In the case of three qubits
  we then showed how to switch on and off genuine tripartite entanglement by slightly moving one of the three qubits.
   For six qubits, we found a large amount of entanglement under different bipartite forms, produced during the
   dissipative dynamics. Finally, by varying the number of qubits from 2 to 10, we showed that stable values of
   entanglement occur in the more strongly correlated bipartitions.

Our analysis pointed out that simple OTE
   configurations may permit the production and manipulation of steady multipartite entanglement, resistant
   for large inter-qubits distances (up to 100 $\mu$m in Fig. \ref{fig3}), offering then new tools possibly
   exploitable for quantum computational tasks. These phenomena could be observed, for example, in configurations
   similar to the one considered in \cite{ObrechtPRL07}, where trapped atoms are placed in proximity to a substrate
    held at a temperature different from that of the cell surrounding the emitters and the substrate. The atomic-like systems of our study could be implemented, for example, with cold atoms trapped near surfaces \cite{Goban14}, possibly in symmetric configurations \cite{Nogrette14}, or with quantum dots.

\begin{acknowledgments} Authors  thank N. Bartolo and B. Leggio for useful discussions and acknowledge financial
support from the Julian Schwinger Foundation.
\end{acknowledgments}

\appendix

 \section{Projection of the master equation}\label{appendix}

Here we project the master equation of Eq. \eqref{master equation} on the basis of eigenstates of  $H_{\mathrm{eff}}/\hbar$, given in Eq. \eqref{M}.

In the following the eigenstates of $H_S/\hbar$ spanning the sector $W^n$  are indicated as
$\ket{\mathbf{k}_\beta^{(n)}}=\ket{\{k_1,...., k_n\}_\beta}$,
being $1\leq k_1 \leq ...\leq k_n\leq N$. For each number of excitations $n$,
 we can span $W^n$ in terms of the right eigenstates of $H_{\mathrm{eff}}/\hbar$:
 \begin{equation}\label{eigenstatesofM}
 (H_{\mathrm{eff}}/\hbar) \ket{\lambda_\alpha^{(n)}}=\Omega_{\alpha}^{(n)} \ket{\lambda_\alpha^{(n)}}= (G_{\alpha}^{(n)}+ i F_{\alpha}^{(n)})\ket{\lambda_\alpha^{(n)}},
 \end{equation}
 where for each $n$, $\alpha$ runs from 1 to $d_n=N!/[(N-n)! n!]$.
Now we write
\begin{equation}
\ket{\lambda_\alpha^{(n)}} = \sum_{\beta} C_{\alpha, \beta}^{(n)}  \ket{\mathbf{k}_\beta^{(n)}},
\end{equation}
where for each $n$, $C^{(n)}$ is the matrix allowing the change of basis.
We now invert the above relationship to get
\begin{equation}
\ket{\mathbf{k}_\beta^{(n)}}= \sum_{\alpha} [C^{(n)}_{\beta,\alpha}]^{-1} \ket{\lambda_\alpha^{(n)}},
\end{equation}
where the matrix $[C^{(n)}]^{-1}$ is the inverse of $C^{(n)}$ (we indicate its elements with $[C^{(n)}_{\beta,\alpha}]^{-1}$).
In order to project the master equation in the basis of the right eigenstates of $ H_{\mathrm{eff}}/\hbar$ we write down
 the following expressions
\begin{equation}\label{actionsigmaim} \begin{split}
  \sigma_i^-\ket{\lambda_\alpha^{(n)}} &= \sum_{\beta \in \Theta_i^{(n)}} C_{\alpha, \beta}^{(n)}
   \ket{\mathbf{k}_{\beta_{-i}}^{(n-1)}}\\ &=\sum_{\bar{\alpha}} \sum_{\beta \in \Theta_i^{(n)}}
    C_{\alpha, \beta}^{(n)}   [C^{(n-1)}_{ \beta_{-i},\bar{\alpha}} ]^{-1}
    \ket{\lambda_{\bar{\alpha}}^{(n-1)}}\\&=\sum_{\bar{\alpha}}
    B^{(n,-i)}_{\alpha, \bar{\alpha}} \ket{\lambda_{\bar{\alpha}}^{(n-1)}},
 \end{split}\end{equation}
where the action of $\sigma_i^-$ is such that in the expression of $\ket{\lambda_\alpha^{(n)}}$ in terms of
$\ket{\mathbf{k}_\beta^{(n)}}$ only the $n$-tuples $\{k_1,...., k_n\}$ (individuated by $\beta$) containing
 $i$ survive [we name this ensemble $ \Theta_i^{(n)}$ and $\beta_{-i}$ the index individuating the eigenvector in
 $W^{(n-1)}$ obtained by applying $\sigma_i^-$ to the eigenvector  in $W^{(n)}$ individuated by $\beta$].
 With $\mathbf{k}_{\beta_{-i}}^{(n-1)} $ we indicate the resulting $(n-1)$-tuple after that the element $i$ is
 erased by the action of  $\sigma_i^-$ (qubit $i$ passes from the excited to the ground state). We have also defined
\begin{equation}
B^{(n,-i)}_{\alpha, \bar{\alpha}} = \sum_{\beta \in \Theta_i^{(n)}} C_{\alpha, \beta}^{(n)}
[C^{(n-1)}_{ \beta_{-i},\bar{\alpha}} ]^{-1}.
\end{equation}

Concerning the action of $\sigma_i^+$, it is
\begin{equation}\label{actionsigmaip}\begin{split}
  \sigma_i^+\ket{\lambda_\alpha^{(n)}} &= \sum_{\beta \in \Theta^{(n)}_{-i}} C_{\alpha, \beta}^{(n)}
   \ket{\mathbf{k}_{\beta_{+i}}^{(n+1)}}\\ &=\sum_{\bar{\alpha}} \sum_{\beta \in \Theta_{-i}^{(n)}}
   C_{\alpha, \beta}^{(n)}   [C^{(n+1)}_{\beta_{+i},\bar{\alpha}} ]^{-1}\ket{\lambda_{\bar{\alpha}}^{(n+1)}}\\&=\sum_{\bar{\alpha}} B^{(n,+i)}_{\alpha, \bar{\alpha}} \ket{\lambda_{\bar{\alpha}}^{(n+1)}},
\end{split}\end{equation}
where the action of $\sigma_i^+$ is such that in the expansion of $\ket{\lambda_\alpha^{(n)}} $ in terms of
$\ket{\mathbf{k}_\beta^{(n)}}$ only the $n$-tuples $\{k_1,...., k_n\}_\beta$ not containing $i$ survive
(that is only if qubit $i $ is in the ground state) [we name this ensemble $ \Theta^{(n)}_{-i}$].
 With $\mathbf{k}_{\beta_{+i}}^{(n+1)}$ we indicate the resulting $(n+1)$-tuple after that the element $i$ is added
 by the action of  $\sigma_i^+$ (qubit $i$ passes from the ground to the excited state). We have also  defined
\begin{equation}
B^{(n,+i)}_{\alpha, \bar{\alpha}}= \sum_{\beta \in \Theta_{-i}^{(n)}} C_{\alpha, \beta}^{(n)}
 [C^{(n+1)}_{\beta_{+i},\bar{\alpha}} ]^{-1}.
\end{equation}

In particular, for the cases $n=0$ and $n=N$ the dimensionality is $d_0=d_N=1$ (the indices $\alpha$ and $\beta$ take only the value 1). It holds
\begin{equation} \label{actionextremes} \begin{split}
\quad \sigma_i^- \ket{\lambda_1^{(N)}}&=   \ket{\mathbf{k}_{1_{-i}}^{(N-1)}}=\sum_{\bar{\alpha}}
 [C^{(N-1)}_{1_{-i},\bar{\alpha}} ]^{-1}\ket{\lambda_{\bar{\alpha}}^{(N -1)}}\\&=
 \sum_{\bar{\alpha}} B^{(N,-i)}_{1, \bar{\alpha}} \ket{\lambda_{\bar{\alpha}}^{(N-1)}}
\\
 \sigma_i^- \ket{\lambda_1^{(0)}}&=0, \\ \sigma_i^+ \ket{\lambda_1^{(N)}}&= 0,\\
 \sigma_i^+\ket{\lambda_1^{(0)}} &= \ket{\mathbf{k}_{1_{+i}}^{(1)}}=\ket{\{i\}}=\sum_{\bar{\alpha}}
  [C^{(1)}_{1_{+i},\bar{\alpha}} ]^{-1}\ket{\lambda_{\bar{\alpha}}^{(1)}}\\&
  =\sum_{\bar{\alpha}} B^{(0,+i)}_{1, \bar{\alpha}} \ket{\lambda_{\bar{\alpha}}^{(n-1)}}.
\end{split}\end{equation}

Using  Eqs. \eqref{master equation 2}, \eqref{actionsigmaim},  \eqref{actionsigmaip} and  \eqref{actionextremes} it is then easy to write the projection
of master equation \eqref{master equation} on the basis of eigenstates of $H_{\mathrm{eff}}/\hbar$:
\begin{equation}\label{dme}\begin{split}
\bra{\lambda_\beta^{(m)}}&\dot{\rho} \ket{\lambda_\alpha^{(n)}}=\dot{\rho}_{\beta_m,\alpha_n} = - i \Big(\Omega_{\beta}^{(m)\,*}-\Omega_{\alpha}^{(n)}\Big) \rho_{\beta_m,\alpha_n} \\ &
+\sum_{\bar{\beta},\bar{\alpha}} \Big[ \sum_{i,j} \Gamma_{ij}^+ B_{\beta,\bar{\beta}}^{(m,+j)\,*} B_{\alpha,\bar{\alpha}}^{(n,+i)}\Big] \rho_{\bar{\beta}_{m+1},\bar{\alpha}_{n+1}}
\\ &
+\sum_{\bar{\beta},\bar{\alpha}}\Big[ \sum_{i,j} \Gamma_{ij}^- B_{\beta,\bar{\beta}}^{(m,-j)\,*} B_{\alpha,\bar{\alpha}}^{(n,-i)}\Big] \rho_{\bar{\beta}_{m-1},\bar{\alpha}_{n-1}}.
\end{split}\end{equation}
We define the new functions
\begin{equation}\label{PandM}\begin{split}
{}_{\alpha_n}^{\beta_m}P_{\bar{\alpha}_{n+1}}^{\bar{\beta}_{m+1}}& =\sum_{i,j} \Gamma_{ij}^+ B_{\beta,\bar{\beta}}^{(m,+j)\,*} B_{\alpha,\bar{\alpha}}^{(n,+i)},
\\ {}_{\alpha_n}^{\beta_m}M_{\bar{\alpha}_{n-1}}^{\bar{\beta}_{m-1}}& = \sum_{i,j} \Gamma_{ij}^- B_{\beta,\bar{\beta}}^{(m,-j)\,*} B_{\alpha,\bar{\alpha}}^{(n,-i)},
\end{split}\end{equation}
and we rewrite Eq. \eqref{dme} as
\begin{equation}\label{dme2}\begin{split}
&\dot{\rho}_{\beta_m,\alpha_n} = - i \Big(\Omega_{\beta}^{(m)\,*}-\Omega_{\alpha}^{(n)}\Big) \rho_{\beta_m,\alpha_n} \\ &
+\sum_{\bar{\beta},\bar{\alpha}} {}_{\alpha_n}^{\beta_m}P_{\bar{\alpha}_{n+1}}^{\bar{\beta}_{m+1}}\rho_{\bar{\beta}_{m+1},\bar{\alpha}_{n+1}}
+\sum_{\bar{\beta},\bar{\alpha}} {}_{\alpha_n}^{\beta_m}M_{\bar{\alpha}_{n-1}}^{\bar{\beta}_{m-1}}\rho_{\bar{\beta}_{m-1},\bar{\alpha}_{n-1}}.
\end{split}\end{equation}
We observe that the choice of the right eigenstates of $H_{\mathrm{eff}}/\hbar$ as basis where to project the master equation leads
to the relevant
simplification that each density matrix element belonging to $W^n$ is not connected to the other elements of  $W^n$
and is only connected to all the populations and coherences of $W^{n+1}$ and $W^{n-1}$.
We finally remark that one can further simplify the problem by using the secular approximation if for each number of excitations $n$
 there are not degenerate eigenstates in the spectrum of $H_{\mathrm{eff}}/\hbar$ \cite{CohenTannoudji97, Freedhoff04}.

\end{document}